# Low-Frequency Wide Band-Gap Elastic/Acoustic Meta-Materials Using the KDamping Concept

**Antoniadis I.A[1], Chatzi E.[2], Chronopoulos D.[3], Paradeisiotis A.[1], Sapountzakis I.[1], S.Konstantopoulos[1]**

[1]Dynamics and Structures, Mechanical Engineering, National Technical University of Athens, Athens, 15780 Zografou, Greece

[2]Structural Mechanics, Structural Engineering, Civil, Environmental and Geomatic Engineering, ETH Zürich, Zürich, Switzerland

[3]Dynamics of Aerospace Structures, Aerospace Technology & Composites Group, Faculty of Engineering, The University of Nottingham, Nottingham, Nottinghamshire, United Kingdom

## Abstract

The terms "acoustic/elastic meta-materials" describe a class of periodic structures with unit cells exhibiting local resonance. This localized resonant structure has been shown to result in negative effective stiffness and/or mass at frequency ranges close to these local resonances. As a result, these structures present unusual wave propagation properties at wavelengths well below the regime corresponding to band-gap generation based on spatial periodicity, (i.e. "Bragg scattering"). Therefore, acoustic/elastic meta-materials can lead to applications, especially suitable in the low-frequency range.

However, low frequency range applications of such meta-materials require very heavy internal moving masses, as well as additional constraints at the amplitudes of the internally oscillating locally resonating structures, which may prohibit their practical implementation.

In order to resolve this disadvantage, the KDamping concept will be analyzed. According to this concept, the acoustic/elastic meta-materials are designed to include negative stiffness elements instead of, or in addition to the internally resonating masses. This concept removes the need for the heavy locally added masses, while it simultaneously exploits the negative stiffness damping phenomenon.

Application of both Bloch's theory and classical modal analysis for the one-dimensional mass-in-mass lattice is analyzed and corresponding dispersion relations are derived. Preliminary results indicate significant advantages over the conventional mass-in-mass lattice, such as broader band-gaps and increased damping ratio and reveal significant potential in seismic meta-structures and low frequency acoustic isolation-damping.

## 1   Introduction

The terms "acoustic/elastic meta-materials" (Hussein et al. 2014) describe a class of periodic structures with unit cells exhibiting local resonance.  An intuitive description of such designs lies in use of a simple "mass-in-mass" lumped parameter model, resulting in negative effective stiffness and/or mass at frequency ranges close to the local resonances. Unlike phononic crystals (Sigalas and Economou 1992), (Vasseur et al. 1994) which are based on Bragg scattering, this localized resonant structure has been shown (Liu et al. 2000), (Huang et al. 2009) to exhibit bandgaps at wavelengths much longer than the lattice size,  thus enabling low-frequency vibration attenuation, wave guiding, and filtering, among other applications. Preliminary applications include among others vibration absorption in beams (Sun et al. 2010), (Xiao et al. 2013) and plates (Pennec et al. 2008), (Peng and Pai 2014).

However, low frequency range applications of such locally resonant meta-materials require very heavy internal moving masses with potentially large response amplitudes of oscillation, which may prohibit their practical implementation.

For example, the relevant applications reported in seismic meta-materials and seismic meta-structures (Cheng and Shi 2014), (Shi et al. 2014), (Huang and Shi 2013), (Colombi et al. 2016), (Palermo et al. 2016) refer to frequencies higher than 2.5Hz with limited band gaps, well above the main frequency range [0.5Hz-3.0Hz] of seismic excitations. Although preliminary results have demonstrated the feasibility of the design of acoustic/elastic meta-material based seismic meta-structures (Dertimanis et al. 2016), (Wagner et al. 2016) with frequency ranges between [0.5Hz-1.5Hz] well within the seismic frequency range, they still require very heavy internal oscillating masses.

Similarly, current applications of locally resonant meta-materials in acoustics (Groby et al. 2014), (Weisser et al. 2016) address frequencies well above 500Hz. No effective applications in low frequency noise isolation with frequencies well below 500Hz have been reported. It should be mentioned, that in this frequency range, the performance of the already existing noise insulation materials is quite poor.

In order to resolve this disadvantage, the paper considers the application of negative stiffness based damping concepts towards the design of highly dissipative low-frequency elastic/acoustic meta-materials. Such an oscillator design concept is proposed in (Antoniadis et al., 2015), incorporating a negative stiffness element, which can exhibit extraordinary damping properties, without presenting the drawbacks of the traditional linear oscillator, or of the 'zero-stiffness' designs. Although the proposed oscillator incorporates a negative stiffness element, it is designed to be both statically and dynamically stable. Once such a system is designed according to the approach proposed in (Antoniadis et al., 2015), it is shown to exhibit an extraordinary damping behaviour.

The above idea of (Antoniadis et al., 2015) is further treated in a systematic way in (Antoniadis et al. 2016), within the context of the design of a general class of tuned mass dampers. The resulting KDamper concept is a novel passive vibration isolation and damping concept, based essentially on the optimal combination of appropriate stiffness elements, which include a negative stiffness element. The KDamper can be considered as an extension of the traditional Tuned Mass Damper (TMD), by the addition of a negative stiffness element to the internal oscillating mass. Contrary to the TMD and its variants, the KDamper substitutes the inertial forces of the added TMD mass by the stiffness force of the negative stiffness element. Among others, this can provide significant comparative advantages, especially in the very low frequency range. As a result, the KDamper can achieve better damping characteristics, without the need of additional heavy masses, as in the case of the TMD. Moreover, since the isolation and damping properties of the KDamper essentially result from the stiffness elements of the system, further technological advantages can emerge, in terms of weight, complexity and reliability.

A preliminary review of the reported physical principles for the design of negative stiffness elements can be found in (Antoniadis et al. 2016). These principles include conventional pre-stressed elastic mechanical elements, such as post-buckled beams, plates, shells and pre-compressed springs arranged in appropriate geometrical configurations, or other sources of forces, such as gravitational, magnetic or electromagnetic.

Based on the periodic repetition of KDamper elements, a novel class of metamaterials can be designed, in the same way that the periodic repetition of TMDs (Sugino et al. 2016) consists the underlying concept of the locally resonant metamaterials. Preliminary results (Chronopoulos et al. 2017), (Chronopoulos et al. 2015) confirm the efficiency of this approach.

Towards this direction, it should be noted that recently, numerous periodic cellular structures with advanced dynamic behavior have been also proposed, (Michelis and Spitas 2010), (Virk et al. 2013), (Baravelli and Ruzzene 2013), (Wang et al. 2014), (Kochmann, 2014), (Harne et al. 2017) combining high positive and negative stiffness elements, such as buckled beams or chiral structures. However, a systematic theoretical framework for the analysis and design of low-frequency wide band-gap elastic/acoustic meta-materials is only presented in this paper.

Application of both Bloch's theory (Section 3) and the classical modal analysis (Section 4) at the one-dimensional mass-in-mass lattice is analyzed and the corresponding dispersion relations are derived. The results (Section 5) indicate significant advantages over the conventional mass-in-a mass lattice, such as broader band-gaps and increased damping ratio and reveal significant potential in the proposed solution. A preliminary feasibility analysis (Section 6) for seismic meta-structures and low frequency elastic/acoustic isolation-damping confirms the strong potential and applicability of this concept.

## 2  Materials & Methods

In order to analyze the 1D lattice resulting from the periodic repetition of KDamper elements, a basic framework needs to be established, regarding the concepts of the KDamper, the negative stiffness and its structural realization.

### 2.1  Overview of the KDamper concept

Figure 1.**C** presents the basic layout of the KDamper concept (Antoniadis et al. 2016), in association to the closely related concepts of the Quasi-Zero Stiffness (QZS) oscillator (Fig.1.**a**), and the Tuned Mass Damper (TMD), Fig.1.**b**. They are all designed to minimize the response $u_S(t)$ of an undamped SDoF system of mass *m* and static stiffness *k* of to an external excitation *f(t)*.

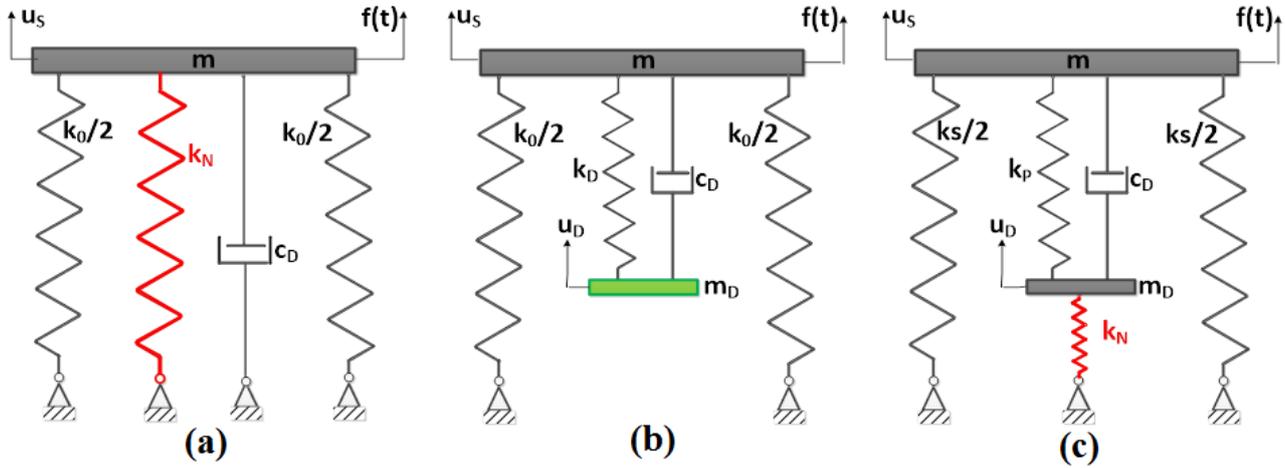

**Figure 1.** Schematic presentation of the considered vibration absorption concepts **(A)** Quasi-Zero Stiffness (QZS) oscillator, **(B)** Tuned Mass Damper (TMD), **(C)** KDamper.

The concept of the QZS oscillator, presented in Fig.1.**a**, is to add a negative stiffness element $k_N$ in parallel to the original stiffness $k_0$ of the system, so that the overall stiffness of the system becomes:

$$k_{QZS} = k_0 + k_N \leq k_0 \qquad (1)$$

However, this limits the static loading capacity of the structure, which may result to unsolvable problems, especially for vertical vibration isolation.

Figure 1.**c** presents the fundamental concept of the KDamper. Similarly to the QZS oscillator, it uses a negative stiffness element $k_N$. However, contrary to the QZS oscillator, the first basic requirement of the KDamper is that the overall static stiffness $k_0$ of the system is maintained:

$$k_0 = k_S + \frac{k_P k_N}{k_P + k_N} \qquad (2)$$

In this way, the KDamper can overcome the fundamental disadvantage of the QZS oscillator. Compared to the TMD, the KDamper uses an additional negative stiffness element $k_N$, which connects the additional mass also to the base. Thus, the equation of motion of the KDamper becomes:

$$m\,\ddot{u}_S + c_D(\dot{u}_S - \dot{u}_D) + k_S u_S + k_P(u_S - u_D) = f(t) \qquad (3.a)$$

$$m_D \ddot{u}_D - c_D(\dot{u}_S - \dot{u}_D) - k_P(u_S - u_D) + k_N u_D = 0 \qquad (3.b)$$

Assuming a harmonic excitation in the form of

$$f(t) = F_0 e^{i\omega t} \qquad (4.a)$$

and a steady state complex response of:

$$u_S(t) = \tilde{U}_S e^{i\omega t} \qquad (4.b)$$

$$u_D(t) = \tilde{U}_D e^{i\omega t} \qquad (4.c)$$

the equations of motion of the Kdamper become:

$$-\omega^2 m \tilde{U}_S + i\omega c_D(\tilde{U}_S - \tilde{U}_D) + k_S \tilde{U}_S + k_P(\tilde{U}_S - \tilde{U}_D) = F_0 \qquad (5.a)$$

$$-\omega^2 m_D \tilde{U}_D - i\omega c_D(\tilde{U}_S - \tilde{U}_D) - k_P(\tilde{U}_S - \tilde{U}_D) + k_N \tilde{U}_D = 0 \qquad (5.b)$$

or equivalently:

$$-\omega^2 m \tilde{U}_S + k_S \tilde{U}_S - \omega^2 m_D \tilde{U}_D + k_N \tilde{U}_D = F_0 \qquad (6.a)$$

$$-\omega^2 m_D \tilde{U}_D - i\omega c_D(\tilde{U}_S - \tilde{U}_D) - k_P(\tilde{U}_S - \tilde{U}_D) + k_N \tilde{U}_D = 0 \qquad (6.b)$$

A careful examination of Eq. (6.a) reveals that the amplitude $F_{MD}$ of the inertia force of the additional mass $m_D$ and the amplitude $F_N$ of the negative stiffness force are exactly in phase, due to the negative value of $k_N$

$$F_{MD} = -\omega^2 m_D \tilde{U}_D \qquad (7.a)$$

$$F_N = k_N \tilde{U}_D \leq 0 \qquad (7.b)$$

Thus, the KDamper essentially consists an indirect approach to increase the inertia effect of the additional mass $m_D$ without however increasing directly the mass $m_D$ itself. Moreover, it should be noted that the value of $F_{MD}$ depends on the frequency, while the value of $F_N$ is constant in the entire frequency range, a fact which is of particular importance in low frequency vibration isolation.

The basic parameters of the KDamper are $\mu, \kappa$ and $\rho$ where

$$\mu = m_D/m \qquad (8)$$

$$\omega = \omega_D/\omega_0 \qquad (9)$$

The optimal value of $\rho$ where the transfer function of the displacement $u_S$ (Fig.**1.c**) with regards to the excitation amplitude is minimized at a selected frequency, is

$$\rho_{OPT} = \sqrt{\frac{1}{(1+\mu+\kappa\mu)(1+\mu) - \kappa^2\mu}} \tag{10}$$

in terms of the parameters $\mu$ and $\kappa$. This design concept is exactly the same, as the one used by DenHartog for the selection of the conventional TMD design parameters. According to this choice of $\rho$ and in view of the denominator of Eq.(10), the maximum value of parameter κ is calculated as:

$$\kappa_{MAX} = (1+\mu)\frac{1+\sqrt{1+4/\mu}}{2} \tag{11}$$

Therefore, the value of $\kappa$ can be selected as a fraction of $\kappa_{MAX}$, namely

$$\kappa = (0:1)\kappa_{MAX} \tag{12}$$

As presented in detail in (Antoniadis et al. 2016), the oscillator presents better vibration absorption capabilities when $\kappa \to \kappa_{MAX}$, but at the same time needs to be considered that the static stability margin tends to zero because then the overall static stiffness of the oscillator [Eq.(2)] becomes negative.

## 2.2 The negative stiffness concept

What is known as positive stiffness, occurs when the deformation is in the same direction as the applied force, corresponding to a restoring force that returns the deformable body to its neutral position. Physically, an elastic object is expected to resist when pressed, by exerting a restoring force. Negative stiffness involves a reversal of the usual directional relationship between force and displacement in deformed objects. In structural mechanics, negative stiffness is obtainable through post-buckling processes, which is a structural effect. In solid materials, a negative curvature in system free energy is predicted via the Landau phenomenological theory on phase transformation, i.e. negative stiffness, in the vicinity of phase transitions. A material with negative stiffness is in unstable equilibrium, because the material has a higher positive stored energy at equilibrium, compared to neighboring possible equilibrium configurations. While negative stiffness is unstable, it can be stabilized by positive stiffness surroundings. Thus, a material with negative stiffness can be stable if it is constrained. In summary, the sources of negative stiffness are pre-load, geometric nonlinear effects or phase transitions.

## 2.3 Indicative realization of a negative stiffness element

The aforementioned structural realization of a negative stiffness element -not in a material level- in this particular case, refers to the use of regular pre-stressed stiffness elements, resulting in a negative stiffness element. An indicative implementation of a negative stiffness element via structural means incorporated in the KDamper, is depicted in Fig.**2**.

The negative stiffness spring $k_N$ is realized by a set of two symmetric horizontal linear springs with constants $k_H$, which support the mass $m_D$ by an articulated mechanism. The static equilibrium position of the system is depicted in Fig.**2.a** under the action of the gravity force. The perturbed position after an external dynamic excitation *f(t)* is depicted in Fig.**2.b** along with the necessary notation concerning the various displacements of the system. The equations of motion of the proposed oscillator are:

$$m\ddot{x} + c_D(\dot{x}-\dot{y}) + k_S(l_S - l_{SI}) + k_P(l_P - l_{PI}) = f + mg \tag{13.a}$$

$$m_D\ddot{y} - c_D(\dot{x}-\dot{y}) - k_P(l_P - l_{PI}) + f_N(u) = m_D g \tag{13.b}$$

where

- $l_S(t)$: length of the spring $k_S$
- $l_{SI}$: initial length of the un-deformed spring $k_S$
- $l_P(t)$: length of the spring $k_P$
- $l_{PI}$ initial length of the undeformed spring $k_P$ and
- $f_N(u)$ non-linear force exerted by the set of the two symmetric oblique springs $k_H$

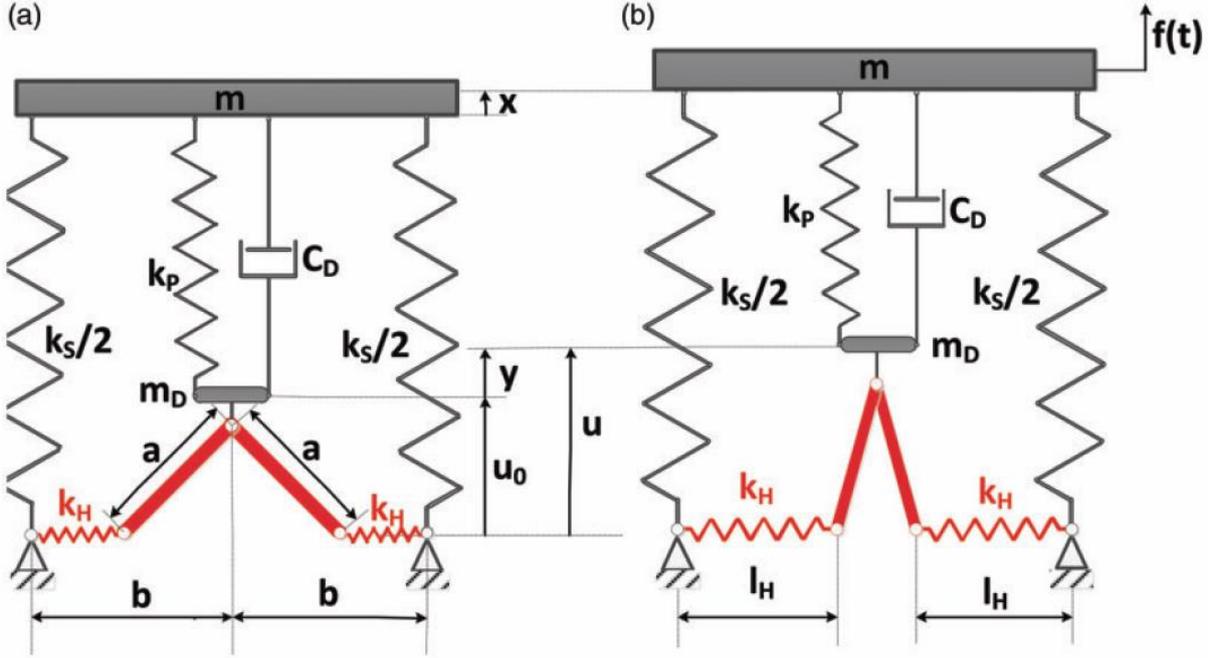

**Figure 2.** Schematic presentation of the realization of the KDamper concept using a non-linear pair of horizontal springs (a) Configuration at the static equilibrium point (b) Notation concerning the perturbed configuration.

Considering the potential energy $U_N$ of the configuration

$$U_N[u(y)] = 2\frac{1}{2}k_H(l_H - l_{HI})^2 \qquad (14)$$

the following expressions for the non-linear force $f_N$ and the equivalent non-linear stiffness $k_N$ of the set of the horizontal springs $k_H$ are derived:

$$f_N(u) = \frac{\partial U_N}{\partial y} = \frac{\partial U_N}{\partial u} = -2k_H\left(1 + c_I\frac{1}{\sqrt{1 - u^2/a^2}}\right)u \qquad (15)$$

$$k_N = \frac{\partial f_N}{\partial y} = \frac{\partial f_N}{\partial u} = -2k_H\left[1 + c_I\frac{1}{(1 - u^2/a^2)^{3/2}}\right] \qquad (16)$$

where

$$l_H = b - \sqrt{a^2 - u^2} \qquad (17)$$

$$c_I = (l_{HI} - b)/a \qquad (18)$$

- $l_{HI}$: the initial length of the un-deformed springs $k_H$
- $l_H$: the length of the springs $k_H$
- $c_I$: a non-dimensional parameter

Indicatively, in the case of $c_I = 0$, the two horizontal springs are equivalent to a spring with a constant negative stiffness $k_N = -2k_H$.

## 2.4 Bloch analysis of KDamper based meta-materials

The KDamper based meta-materials essentially comprise a periodic repetition of unit lattice cells of KDamper elements, as presented in Fig. 3.

As evident from Fig.3 after setting $k_N=0$, the KDamper concept forms the fundamental underlying concept of the KD meta-materials, in the same way that the TMD forms the underlying concept of acoustic/elastic locally resonant meta-materials.

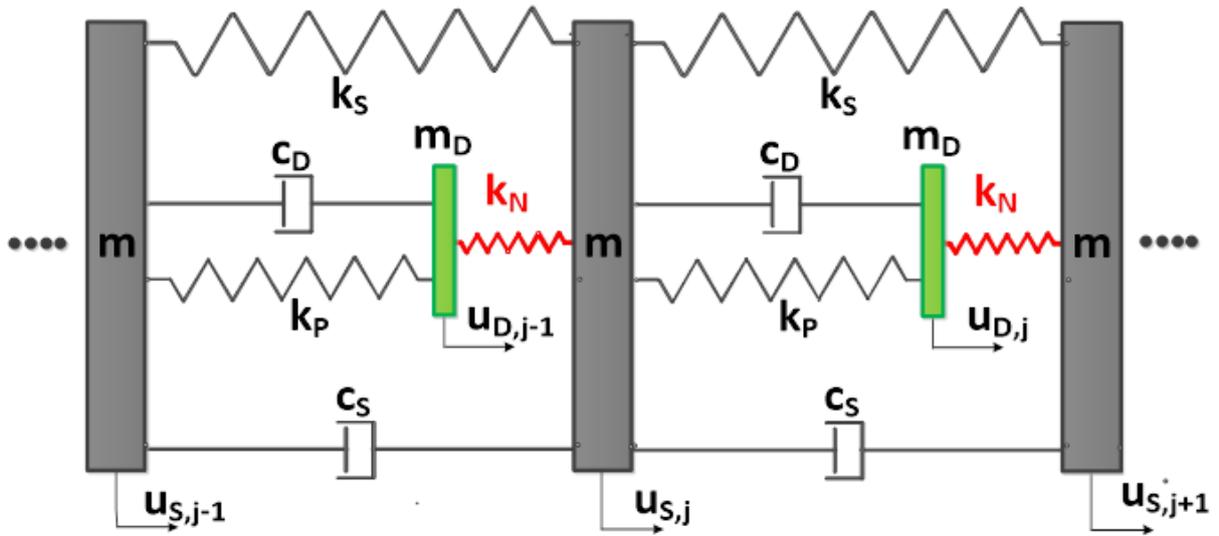

**Figure 3**. Schematic presentation of KDamper meta-materials.

In view of Fig.3, the equations of motion of a typical unit cell becomes:

$$m\,u_{S,j} + c_S(u_{s,j} - u_{s,j-1}) + c_S(u_{s,j} - u_{s,j+1}) + k_S(u_{s,j} - u_{s,j-1}) + k_S(u_{s,j} - u_{s,j+1}) \qquad (19.a)$$
$$+ c_D(u_{s,j} - u_{D,j}) + k_P(u_{s,j} - u_{D,j}) + k_N(u_{s,j} - u_{D,j-1}) = 0$$

$$m_D u_{D,j} + c_D(u_{D,j} - u_{s,j}) + k_P(u_{D,j} - u_{s,j}) + k_N(u_{D,j} - u_{s,j+1}) = 0 \qquad (19.b)$$

Application of the generalized form (Hussein and Frazier 2013) of Bloch's theorem as presented in Section 5.1 of the Appendix, yields a characteristic equation of the form:

$$a_4\lambda^4 + \alpha_3\lambda^3 + \alpha_2\lambda^2 + \alpha_1\lambda + \alpha_0 = 0 \qquad (20)$$

where the coefficients $\alpha_4$, $\alpha_3$, $\alpha_2$, $\alpha_1$ and $\alpha_0$ are also defined in Section 5.1 of the Appendix. The roots obtained from Eq.(20) may be expressed as:

$$\lambda_B = -\zeta_B(k)\omega_B \pm i\omega_B\sqrt{1-\zeta_B^2} \;,\; B = 1,2 \tag{21}$$

where $B$=1 represents the lower branch number, $B$=2 represents the upper branch number of the two dispersion curves and $\omega_B$, $\zeta_B$ represent their corresponding natural frequency and damping ratio.

In order to obtain the band-gap limits, from the resulting Eq.(20) of the Bloch analysis an undamped system is assumed ($\alpha_3=\alpha_1=0$, $\lambda_B=\pm j\omega_B$), which leads to:

$$\alpha_4\omega^4 - (\alpha_{20} + \gamma\alpha_{2\gamma})\omega^2 + \gamma\alpha_{0\gamma} = 0 \tag{22}$$

where:

$$\alpha_{20} = (m_D + m)(k_P + k_N) = (m_D + m)k_I \tag{23.a}$$

$$a_{2\gamma} = k_S m_D \tag{23.b}$$

$$\alpha_{0\gamma} = k_S(k_P + k_N) + k_P k_N = k_0 k_I \tag{23,c}$$

$$k_I = k_P + k_N \tag{24}$$

Solving Eq.(22) for the parameter $\gamma$, yields:

$$\gamma = \omega^2 \frac{\alpha_{20} - \alpha_4\omega^2}{\alpha_{0\gamma} - \alpha_{2\gamma}\omega^2} \tag{25.a}$$

$$\cos(kl) = \cos q = 1 - \frac{\omega^2}{2}\frac{\omega_H^2\omega_I^2 - \omega^2}{\omega_0^2\omega_I^2 - \omega_S^2\omega^2} \tag{25.b}$$

where

$$\omega_0^2 = k_0/m \tag{26.a}$$

$$\omega_I^2 = k_I/m_D \tag{26.b}$$

$$\omega_H^2 = \omega_I\sqrt{1+\mu} \tag{26.c}$$

$$\omega_S^2 = k_S/m \tag{26.d}$$

In view of the denominator and nominator of Eq.(25.b), the band-gap limits are (Huang and Sun 2009):

$$\omega_L \leq \omega \leq \omega_H \tag{27}$$

where:

$$\omega_L^2 = \omega_I \frac{\omega_0}{\omega_S} = \frac{\omega_0^2}{1/\rho^2 + \kappa(1+\kappa)\mu} = \frac{\omega_I^2}{1 + \kappa(1+\kappa)\mu\rho^2} \tag{28.a}$$

$$b_W = \frac{\omega_H - \omega_L}{\omega_L} = \sqrt{(1+\mu)[1 + \kappa(1+\kappa)\mu\rho^2]} - 1 \tag{28.b}$$

$$\kappa = -k_N/k_I \tag{29.a}$$

$$\rho = \omega_I/\omega_0 \tag{29.b}$$

By setting the value of $\kappa$ equal to zero, the corresponding band-gap limits of the acoustic meta-material are obtained from Eqs. (28):

$$\omega_L \xrightarrow[\kappa \to 0]{} \omega_I = \omega_D = \sqrt{k_P/m_D} \tag{30.a}$$

$$b_W \xrightarrow[\kappa \to 0]{} \sqrt{1+\mu} - 1 \tag{30.b}$$

As it can be observed from Eqs. (28), the normalized band gap width $b_W$ can be increased not only by increasing the parameter $\mu$ (i.e., the value of the internally oscillating mass $m_D$), but also the value of the parameter $\kappa$ (i.e., the value of the negative stiffness element $k_N$). This can be considered among others as a consequence of Eq.(7), since the usage of a negative stiffness element can be considered as an indirect approach to artificially increase the inertia of the system.

However, if the value of $\kappa$ increases beyond the upper bound $\kappa_{max}$, the overall static stiffness $k_0$ of the system in Eq.(2) becomes negative, and thus the system is rendered statically unstable.

## 2.5 A modal analysis approach for periodic KDamper metamaterials

An alternative equivalent approach for band-gap estimation is proposed in (Sugino et al. 2016). Since this approach is based on modal analysis, it provides a better insight for engineers to the dynamics of elastic/acoustic meta-materials, it relaxes the assumption of Bloch analysis for a perfect spatial repetition and in any case, it can be used for an independent validation of the results of Bloch analysis.

In order to apply this approach, the harmonic response of the undamped system is considered, with Eqs. (19) leading to:

$$\begin{aligned}-\omega^2 m U_{S,j} + k_S(U_{S,j} - U_{S,j-1}) + k_S(U_{S,j} - U_{S,j+1}) \\ + k_P(U_{S,j} - U_{D,j}) + k_N(U_{S,j} - U_{D,j-1}) = 0\end{aligned} \tag{31.a}$$

LOW-FREQUENCY WIDE BAND-GAP METAMATERIALS USING THE K-DAMPING CONCEPT

$$-\omega^2 m_D U_{D,j} + k_P(U_{D,j} - U_{S,j}) + k_N(U_{D,j} - U_{S,j+1}) = 0 \tag{31.b}$$

The application of this approach in this particular case, which is presented in detail in Section 5.2 of the Appendix, leads to the following set of decoupled equations in non-dimensional form:

$$\Omega_k^2(\Omega_L^2 - \Omega^2) - \Omega^2(\Omega_H^2 - \Omega^2) = 0 \tag{32.a}$$

therefore

$$\Omega^4 - \Omega^2(\Omega_k^2 + \Omega_H^2) + \Omega_k^2\Omega_L^2 = 0 \tag{32.b}$$

where:

$$\Omega = \omega/\omega_I \tag{33.a}$$

$$\Omega_k = \omega_k/\omega_I \tag{33.b}$$

$$\Omega_L = \omega_L/\omega_I \tag{33.c}$$

$$\Omega_H = \omega_H/\omega_I \tag{33.d}$$

Equation (32.b) can have two real roots:

$$\Omega_{MIN}^2 = \frac{\Omega_k^2 + \Omega_H^2}{2}\left[1 - \sqrt{1 - \frac{4\Omega_K^2\Omega_L^2}{(\Omega_k^2 + \Omega_H^2)^2}}\right] \tag{34.a}$$

$$\Omega_{MAX}^2 = \frac{\Omega_k^2 + \Omega_H^2}{2}\left[1 + \sqrt{1 - \frac{4\Omega_K^2\Omega_L^2}{(\Omega_k^2 + \Omega_H^2)^2}}\right] \tag{34.b}$$

The band-gap limits [$\Omega_L,\Omega_H$] of the system result (Sugino et al. 2016) from Eqs. (34) as:

$$\Omega_{MIN}^2 \xrightarrow[\Omega_k \to \infty]{} \Omega_L \tag{35.a}$$

$$\Omega_{MAX}^2 \xrightarrow[\Omega_k \to \infty]{} \Omega_H \tag{35.b}$$

## 3 Results

### 3.1 Band-gap width

The graphical represantion of Eq.(28.b) in Fig. 4 shows the resulting normalized band gap width $b_W$ for a simultaneous variation of both the $\kappa,\mu$ parameters of the KDamper metamaterials. As it can be observed, a bandgap of $b_W=3$ may be already achieved for values of $\kappa$ close to $\kappa_{MAX}$, even for negligible values of the internal oscillating mass ($\mu\approx0$) . Bandgaps of such a width may not be accomplished by conventional acoustic meta-materials, even for extremely high values of the internal oscillating mass ($\mu\geq10$).

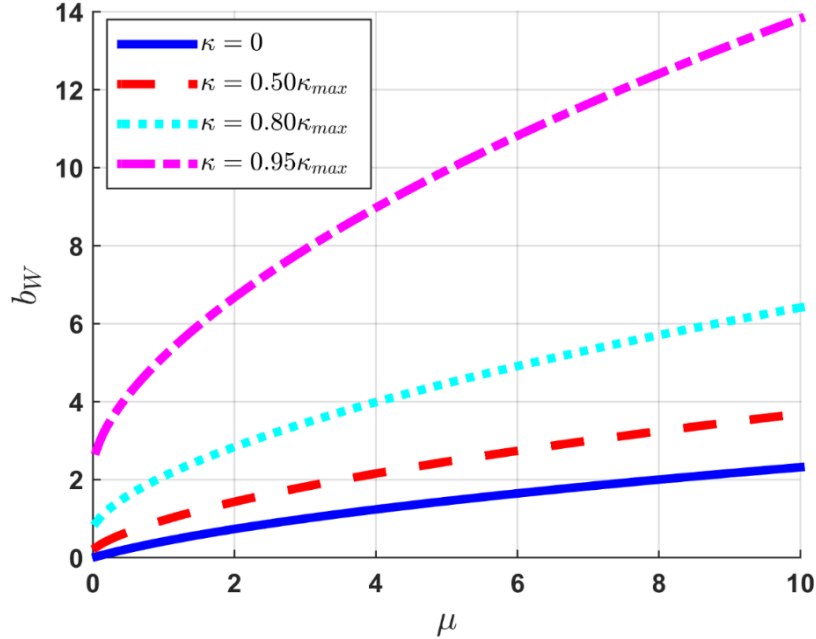

**Figure 4**: Effect of the KDamper design parameters $\kappa,\mu$ on the band-gap width $b_W$

It should be noted, that in view of Eqs. (28), a third parameter $\rho$ in Eq.(29.b) is affecting the bandgap width $b_W$. The value of $\rho$ in Fig.4 has been chosen as in Eq.(10), according to the traditional Kdamper design concept (Antoniadis et al. 2016). According to this choice of $\rho$ the maximum value of parameter $\kappa$ is calculated from Eq.(12). However, in the case of KDamper based metamaterials, the value of $\rho$ can consist a third, freely selectable design parameter, additional to $\kappa$ and $\mu$.

## 3.2 Application examples

Although the KDamper concept can find multiple applications in low-frequency damping and absorption, two critical application ranges are addressed in this paper:

### 3.2.1 Seismic meta-structures

Preliminary results in (Dertimanis et al. 2016), (Wagner et al. 2016), have demonstrated the feasibility of the application of acoustic meta-material concepts for the design of seismic protection zones, consisting of pile arrays. The bandgap achieved is [0.5Hz – 1.5Hz]. The seismic meta-structures implemented have relevant unit lattice parameters as per the first row of Table 1. A number of $M=8$ piles has been demonstrated as sufficient in order to achieve very good seismic wave attenuation properties. However, an internal oscillating mass of $m_D=3021Kg$ is required, which is nine times higher than the external mass of $m=378Kg$.

An alternative design, based on the KDamper meta-material concept is presented in the second row of Table 1. An inclusion of a negative stiffness element, together with a redistribution of the stiffness of the remaining elastic elements of the system can lead to exactly the same band-gap -noted with black dashed lines in Fig.5- however via use of a nine times lighter internal oscillating mass.

**Table 1:** Unit lattice parameters of the seismic meta-structures considered in Section 6.1

|  | m (Kg) | $m_D$ (Kg) | $k_S$ (N/m) | $k_P$ (N/m) | $k_N$ (N/m) | $c_S$ (Ns/m) | $c_D$ (Ns/m) |
|---|---|---|---|---|---|---|---|
| **Acoustic Metamaterial** | 378 | 3021 | 298176 | 29818 | 0 | 212.33 | 189.32 |
| **KDamper metamaterial** | 378 | 378 | 129510 | 50365 | -33576 | 212.33 | 189.32 |

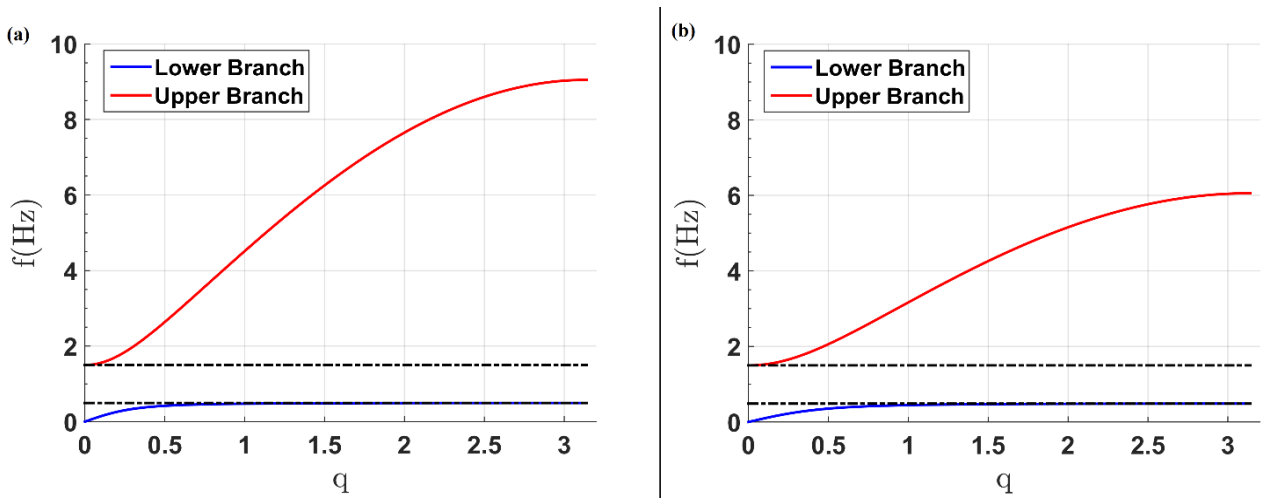

**Figure 5:** Frequency band-structure of the **(A)** Acoustic and the **(B)** KDamper seismic meta-structures

Moreover, the damping ratios (Fig.6) achieved via the KDamper meta-material are significantly higher than those of the conventional acoustic meta-material. This is perhaps of greater significance, since the necessary number of unit cells in order to achieve signifficant seismic wave attenuation can become signifficanlty less than the $M$=8 cells required by the acoustic metamaterial.

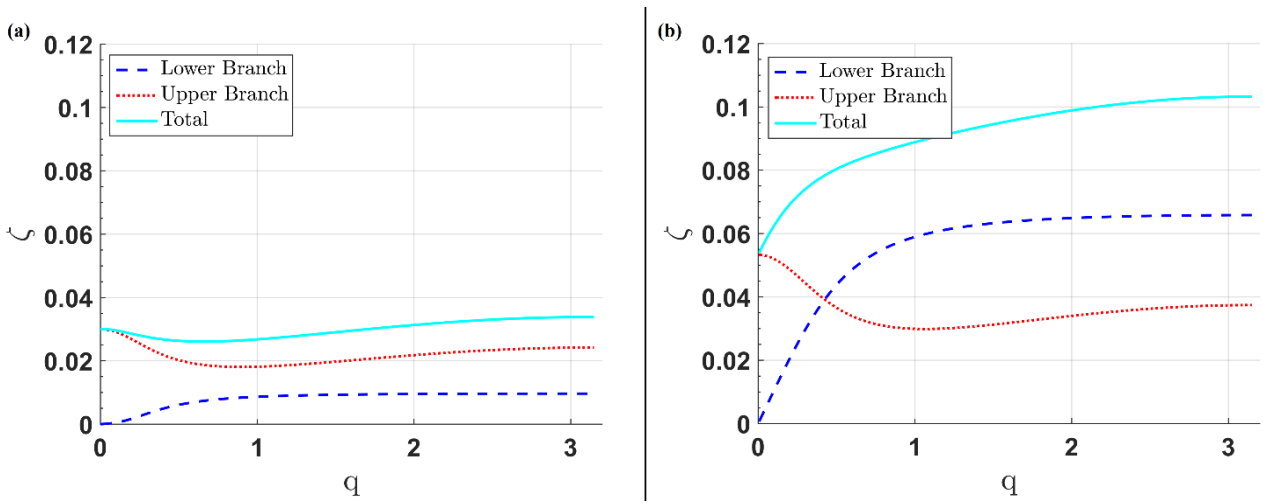

**Figure 6:** Damping ratio band-structure of **(A)** the Acoustic and the **(B)** KDamper seismic meta-structures

### 3.2.2 Middle and low frequency elastic/acoustic meta-materials

Figure 7 presents an indicative concept for the implementation of KDamper based meta-materials towards the design of acoustic isolation-absorption panels.

A conventional "sandwich type" noise isolation panel, consisting of two plates with a noise insulation material between them, can be considered as a simple mass-spring-mass ("*m-k-m*") system, with some degree of internal damping. In view of Fig. 7, this corresponds to a system with $M=1$, $m_D=c_D=k_N=k_P=0$. Periodic repetition of this lattice results to a primitive form of a "phononic crystal" type acoustic insulation meta-material.

The introduction of an internal structure between the masses in the form of a TMD ($m_D$, $c_D$, $k_P \neq 0$) results to an acoustic meta-material design. The additional inclusion of a negative stiffness element ($k_N \neq 0$), results to a KDamper based meta-material.

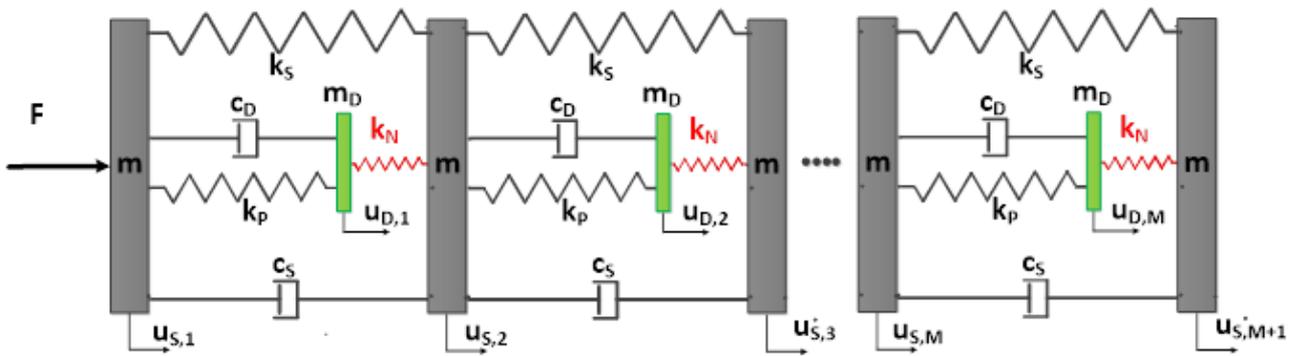

**Figure 7:** The KDamper concept for noise insulation-absorption

The isolation-absorption properties of these three alternative concepts are then compared. The various meta-structures implemented have relevant unit lattice parameters as per Table 2.

**Table 2:** Unit lattice parameters of the elastic/acoustic meta-structures considered in Section 6.2

|  | **m (Kg)** | **$m_D$ (Kg)** | **$k_S$ (N/m)** | **$k_P$ (N/m)** | **$k_N$ (N/m)** | **$c_S$ (Ns/m)** | **$c_D$ (Ns/m)** |
|---|---|---|---|---|---|---|---|
| **Periodic mass-spring-mass system** | 1.00 | 0 | 2467400 | 0 | 0 | 942.48 | 0 |
| **Acoustic Metamaterial** | 1.00 | 0.50 | 2467400 | 789570 | 0 | 942.48 | 888.44 |
| **KDamper metamaterial** | 1.00 | 0.01 | 15890000 | 1221000 | -1119200 | 942.48 | 45.10 |

The selection of the parameters ensures that all three unit lattice types have the same overall static stiffness. For this last reason, the overall static stiffness $k_S$ of the periodic mass-spring-mass system is the same as the static stiffness of the elastic/acoustic metamaterial. However, in view of Eq (2), the value of $k_S$ has to be increased for the KDamper metamaterial, so that its overall static stiffness (now $k_0$) is equal to the overall static stiffness $k_S$ of the other two types.

The Acoustic Metamaterial is designed with a parameter *μ=0.50*. The parameters of the KDamper metamaterial are selected as *μ=0.01, κ=11*. In both cases, the same low band gap frequency of $f_L$=200Hz is selected.

The effect of the three different approaches on the transfer function of the system are presented in Figure 8 using for a different number *M* of unit lattices.

The "*m-k-m*" system essentially acts as a low pass filter for the noise with a cutoff frequency of $f_0$=500Hz. The repetition of unit cells does not improve the low pass behavior. It just increases the high frequency filtering capability of this system. This behavior is typical of the current noise isolation panels.

The acoustic metamaterial retains this low pass filter behavior, while it simultaneously introduces a deep but very narrow band gap. The low frequency of this bandgap is $f_D$=$f_L$=200Hz, while its bandwidth is [200Hz-245Hz]. However, in order to achieve this bandgap, an additional mass of 50% of the original mass of the metamaterial is required. The repetition of unit cells does not improve the low pass behavior or the band-gap boundaries. It just increases the high frequency filtering capability of this system, as well as the depth of the bandgap.

The KDamper metamaterial essentially operates as a band-stop filter. Although its high frequency behavior is less satisfactory than the periodic "*m-k-m*" system and the acoustic meta-material, it is very effective in introducing a very wide band gap with a low frequency of $f_L$=200Hz and a bandwidth of [200Hz-510Hz] with a value of only 1% of the original mass of the system. The repetition of unit cells, as further presented in Fig. 9, has increased the depth of the band-gap, and bears a marginal effect on the high frequency behavior of the system.

# 4    Discussion

Although acoustic metamaterials have been long considered to present the only available direction to create bandgaps at wavelengths much longer than the lattice size, and thus enable low-frequency vibration attenuation, the width of the resulting band gaps is very narrow, since it depends on the ratio of the internal oscillating mass to the external mass of the unit lattice. Metamaterials based on the KDamper concept can overcome this disadvantage, since they exploit the additional synergy of a negative stiffness element inclusion. Thus, they prove highly efficient in generating wide bandgaps in low frequency applications, with only a small fraction of the external mass. Appropriate technological implementations of this concept can lead to drastic improvements in all types of low-frequency technological applications, with emphasis in seismic meta-structures and low-frequency noise isolation-absorption.

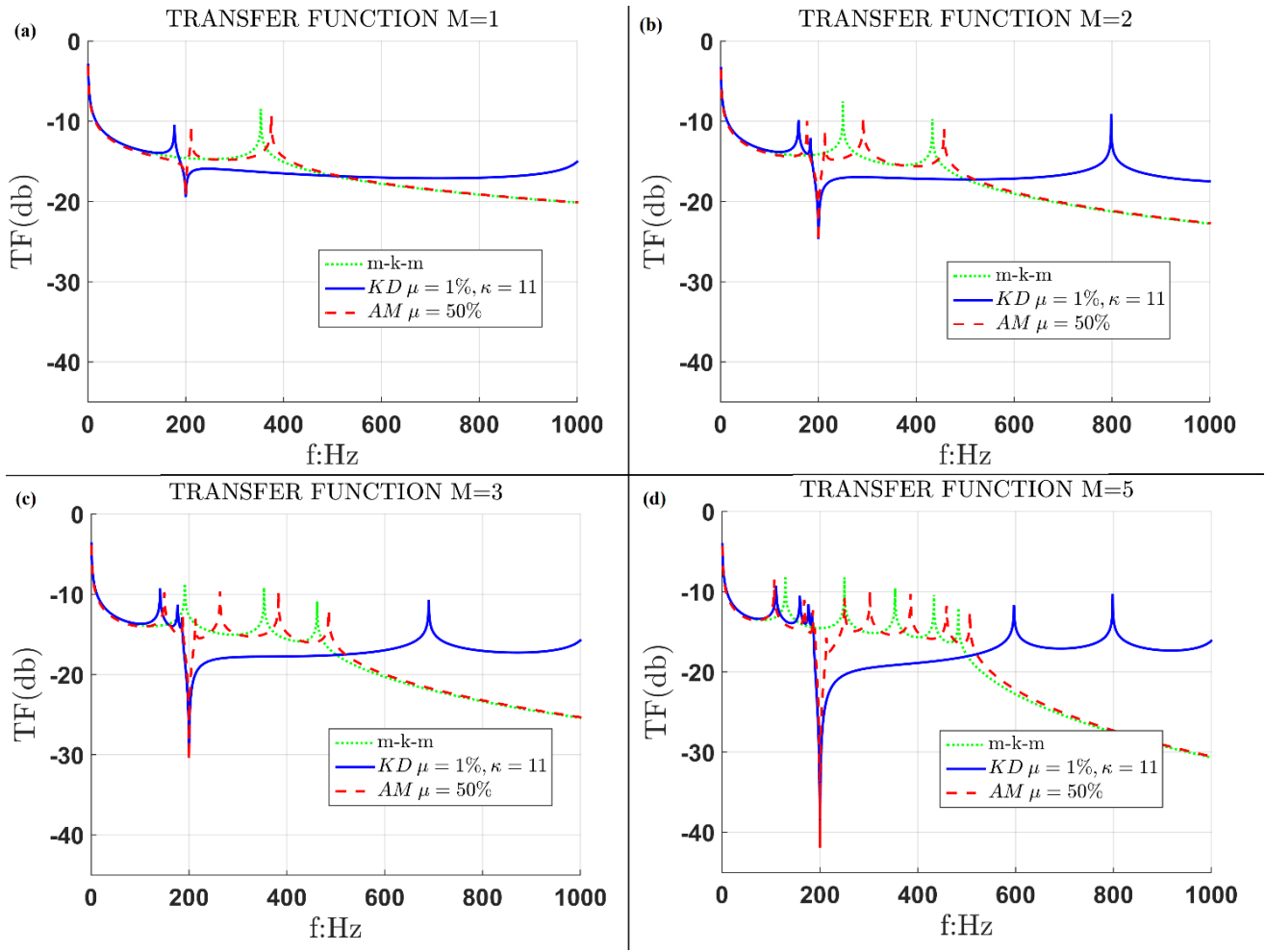

**Figure 8:** Transfer functions of acoustic isolation-absorption systems for number of unit cells **(A)** *M*=1, **(B)** *M*=2, **(C)** *M*=3 and **(D)** *M*=5.

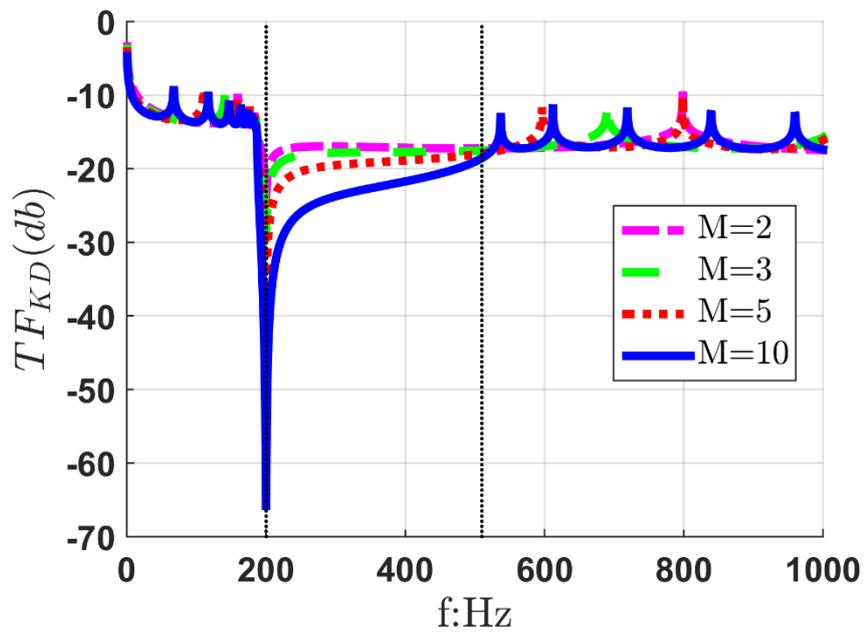

**Figure 9:** Transfer functions of the KDamper metamaterial for various numbers *M* of unit cells.

## 5 Appendix

### 5.1 Application of Bloch's theorem

For each of the system of Eqs. (19.a) and (19.b), the generalized form of Bloch's theorem is applied:

$$u_{S,j} = U_S \, e^{\lambda t} \tag{A1.a}$$

$$u_{S,j+1} = U_S \, e^{ikl+\lambda t} = U_{S,j+1} \, e^{\lambda t} \tag{A1.b}$$

$$u_{S,j-1} = U_S \, e^{-ikl+\lambda t} = U_{S,j-1} \, e^{\lambda t} \tag{A1.c}$$

$$u_{D,j} = U_D \, e^{\lambda t} \tag{A2.a}$$

$$u_{D,j+1} = U_D \, e^{ikl+\lambda t} = U_{D,j+1} \, e^{\lambda t} \tag{A2.b}$$

$$u_{D,j-1} = U_D \, e^{-ikl+\lambda t} = U_{D,j-1} \, e^{\lambda t} \tag{A2.c}$$

where $U_S$, $U_D$ are the wave amplitudes at node $j$, $l$ is the unit-cell length, $k$ is the wave number and $\lambda$ is a complex frequency function that permits wave attenuation in time. In the limiting case of no damping, $\lambda = i\omega$, and the usual form of Bloch's theorem is recovered. Substitution of Eqs. (A1) and (A2) into Eqs. (19) leads to:

$$[\lambda^2 m + \lambda(\gamma c_S + c_D) + (\gamma k_S + k_P + k_N)]U_S - (\lambda c_D + k_P + k_N \, e^{-ikl})U_D = 0 \tag{A3.a}$$

$$[\lambda^2 m_D + \lambda c_D + k_P + k_N]U_D - (\lambda c_D + k_P + k_N \, e^{ikl})U_S = 0 \tag{A3.b}$$

where:

$$\gamma = 2 - (e^{ikl} + e^{-ikl}) = 2[1 - \cos(kl)] = 2(1 - \cos q) \tag{A4.c}$$

$$q = kl \tag{A4.d}$$

Substituting $U_D$ from Eq.(A3.b) into Eq.(A3.a), leads to

$$[a_4 \lambda^4 + \alpha_3 \lambda^3 + \alpha_2 \lambda^2 + \alpha_1 \lambda + \alpha_0 = 0]U_S = 0 \tag{A5}$$

where:

$$a_4 = m_D m \tag{A6.a}$$

$$a_3 = (m_D + m)c_D + \gamma c_S m_D \tag{A6.b}$$

$$a_2 = (m_D + m)(k_P + k_N) + \gamma(c_S c_D + k_S m_D) \tag{A6.c}$$

$$a_1 = \gamma c_S(k_P + k_N) + \gamma c_D(k_S + k_N) \tag{A6.d}$$

$$a_0 = \gamma k_S(k_P + k_N) + \gamma k_P k_N \tag{A6.e}$$

### 5.2 Modal analysis for band-gap estimation

Algebraic elaboration of Eqs.(31) leads to the succession of the following equivalent Eqs. (A7),(A8),(A9):

$$-\omega^2 m U_{S,j} + k_S(2U_{S,j} - U_{S,j-1} - U_{S,j+1}) + (k_P + k_N)U_{S,j} \tag{A7}$$
$$- \left[\frac{k_P^2 U_{S,j} + k_P k_N U_{S,j+1}}{-\omega^2 m_D + k_P + k_N} + \frac{k_N^2 U_{S,j} + k_P k_N U_{S,j-1}}{-\omega^2 m_D + k_P + k_N}\right] = 0$$

$$-\omega^2 \left[m + \frac{m_D(k_P + k_N)}{-\omega^2 m_D + k_P + k_N}\right] U_{S,j} \tag{A8}$$
$$+ \left[k_S + \frac{k_P k_N}{-\omega^2 m_D + k_P + k_N}\right](2U_{S,j} - U_{S,j-1} - U_{S,j+1}) = 0$$

$$-\omega^2 m_{eff} U_{S,j} + k_{eff}(2U_{S,j} - U_{S,j-1} - U_{S,j+1}) = 0 \tag{A9}$$

where:

$$m_{eff} = m + \frac{m_D(k_P + k_N)}{-\omega^2 m_D + k_P + k_N} = m\left[\frac{1 + \mu\omega_I^2}{\omega_I^2 - \omega^2}\right] = m\frac{\omega_H^2 - \omega^2}{\omega_I^2 - \omega^2} \tag{A10}$$

$$k_{eff} = k_S\left[1 + \frac{(k_0/k_S - 1)k_I}{-\omega^2 m_D + k_I}\right] = k_S\left[1 + \frac{(\omega_0^2/\omega_S^2 - 1)\omega_I^2}{\omega_I^2 - \omega^2}\right] \tag{A11}$$
$$= k_S \frac{\omega_L^2 - \omega^2}{\omega_I^2 - \omega^2}$$

and assuming a periodic system without the KDamper elements ($k_P=k_N=c_D=m_D=0$), Eq.(A9) can be written in the following compact matrix notation.

$$-\omega^2 \boldsymbol{M} \boldsymbol{U}_S + \boldsymbol{K}_S \boldsymbol{U}_S = 0 \tag{A12}$$

Using $\omega_k$ and $\boldsymbol{\Phi}_k$ ($k=1, N$) to denote the natural frequencies and the eigenvectors of the system (Eq.(A12)) respectively, and exploiting their orthogonality properties:

$$\boldsymbol{\Phi}_i^T \boldsymbol{M} \boldsymbol{\Phi}_k = 1 \tag{A13.a}$$

$$\boldsymbol{\Phi}_i^T \boldsymbol{K}_S \boldsymbol{\Phi}_k = \omega_k^2 \tag{A13.b}$$

The following modal transform is obtained:

$$\boldsymbol{U}_S = \sum_{k=1}^{N} \boldsymbol{\Phi}_k \eta_k \tag{A13.c}$$

Application of Eqs. (A13) to the system of Eqs. (A9) leads to a set of decoupled equations:

$$-\omega^2(\omega_H^2 - \omega^2)\eta_k - \omega_k^2(\omega_L^2 - \omega^2)\eta_k = 0 \tag{A14}$$